\documentclass[aps,preprint,pra,showpacs]{revtex4}
\usepackage{amsmath,graphicx,bbm,mathrsfs,amssymb,pst-all,bm,color}

\begin{document}

\title{A quantum algorithm for obtaining the energy spectrum of a physical
system without guessing its eigenstates}
\author{Hefeng Wang\footnote{Correspondence to wanghf@mail.xjtu.edu.cn}}
\affiliation{Department of Applied Physics, Xi'an Jiaotong University, Xi'an 710049, China}

\begin{abstract}
We present a quantum algorithm that provides a general approach for
obtaining the energy spectrum of a physical system without making a guess on
its eigenstates. In this algorithm, a probe qubit is coupled to a quantum
register $R$ which consists of one ancilla qubit and a $n$-qubit register
that represents the system. $R$ is prepared in a general reference state,
and a general excitation operator acts on $R$ is constructed. The probe
exhibits a dynamical response only when it is resonant with a transition
from the reference state to an excited state of $R$ which contains the
eigenstates of the system. By varying the probe's frequency, the energy
spectrum and the eigenstates of the system can be obtained.
\end{abstract}

\maketitle

\section{Introduction}

A fundamental problem in the field of many-body systems is to find efficient
ways of simulating Schr\"{o}dinger equations. The main difficulty is that
the dimension of the Hilbert space describing a system of $n$-particles
scales exponentially with $n$. This makes direct numerical simulation of a
large system intractable. On a quantum computer, however, the number of
qubits required to simulate the system increases linearly with the size of
the system. And such problems can be solved efficiently on a quantum
computer.

In quantum chemistry and computational physics, one often has to diagonalize
a large Hamiltonian matrix to obtain the desired eigenvectors and
eigenvalues of a system. Classically, the Davidson's algorithm~\cite%
{davidson} is a large-scale, iterative method which is particularly
effective for extracting selected eigenvectors of a Hamiltonian matrix. In
this algorithm, one has to use a trial wave function. For large systems,
however, this method is expensive and may suffer from slow convergence.
Usually this is due to the fact that the trial wave function is not a good
approximation to the eigenvector of the Hamiltonian matrix. And it is
difficult to find a good trial wave function for a large complicated system,
especially when describing excited states.

In quantum computation, the phase estimation algorithm~(PEA) can be used for
obtaining the eigenvalues of a system~\cite{abrams}. In this algorithm, one
guesses an approximated eigenstate of the system, and prepares this guess
state as input for the algorithm on a quantum computer. The success
probability of the PEA depends on the overlap of the guess state with the
real eigenstate of the system. However, in some cases such as the
bond-dissociation process in chemistry or that related to excited states of
a system, it can become impossible to have a guess state that has any
substantial overlap with the desired eigenstates~\cite{whf}. For many
complicated systems, it is very difficult to make even qualitatively correct
guess on their eigenstates, or to prepare such states on a quantum computer
efficiently.

In Ref.~\cite{wan}, we proposed another quantum algorithm for obtaining the
energy spectrum of a physical system. In this algorithm, one also has to
make guess on the energy eigenstates of the system. The guess state does not
need to have large overlap with any particular eigenstate, but it must have
large overlap with one of the eigenstates of the system in order to achieve
high efficiency.

\section{The algorithm}

In this paper, we present a quantum algorithm for obtaining the energy
spectrum of a physical system that we have no knowledge about its
eigenstates. This algorithm has the following advantages: $\left( i\right) $%
~one does not need to make a guess on any energy eigenstates of the system; $%
\left( ii\right) $~several adjustable elements~(evolution time and
system-probe coupling strength) can be varied to improve the efficiency and
accuracy of the algorithm; $\left( iii\right) $~by introducing a general
reference state and a general excitation operator that can be applied for
any arbitrary physical system, this algorithm provides a general approach
for obtaining the energy spectrum and eigenstates of a system. The details
of the algorithm are shown below.

We let a probe qubit couple to a $(n+1)$-qubit quantum register $R$, which
contains one ancilla qubit and a $n$-qubit quantum register that represents
a physical system of dimension $N=2^{n}$. The Hamiltonian of the whole
system is constructed in the form
\begin{equation}
H=\frac{1}{2}\omega \sigma _{z}\otimes I_{2}^{\otimes \left( n+1\right)
}+I_{2}\otimes \widetilde{H}+c\sigma _{x}\otimes A,
\end{equation}%
where $I_{2}$ is the two-dimensional identity operator. The first term in
the above equation is the Hamiltonian of the probe qubit, the second term is
the Hamiltonian of the register $R$, and the third term describes the
interaction between the probe qubit and $R$. Here, $\omega $ is the
frequency of the probe qubit~($\hbar =1$), and $c$ is the coupling strength
between the probe qubit and $R$, whereas $\sigma _{x}$ and $\sigma _{z}$ are
the Pauli matrices. The Hamiltonian of the register $R$ is in the form
\begin{equation}
\widetilde{H}=\alpha |0\rangle \langle 0|\otimes I_{N}+|1\rangle \langle
1|\otimes H_{S}
\end{equation}%
where $I_{N}$ is the $N$-dimensional identity operator; $\alpha $ is a
parameter that is set as a reference point for the eigenenergy of the
system; $H_{S}$ is the Hamiltonian of the system with dimension of $N$. The
operator $A$ is defined as:
\begin{equation}
A=\sigma _{x}\otimes \left[ \frac{1}{\sqrt{2}}(I_{2}+\sigma _{x})\right]
^{\otimes n}.
\end{equation}%
It acts on the state space of $R$ and plays the role of an excitation
operator.

To run the algorithm, first we prepare the probe qubit in its excited state $%
|1\rangle $ and the register $R$ in a reference state
\begin{equation}
|\Psi _{0}\rangle =\frac{1}{\sqrt{N}}\sum_{j=1}^{N}|\varphi _{j}\rangle
=|0\rangle \otimes \left( \frac{1}{\sqrt{N}}\sum_{j=1}^{N}|j-1\rangle
\right),
\end{equation}%
where $|j-1\rangle $ are the computational basis. This is achieved by
initializing $R$ in state $|0\rangle ^{\otimes \left( n+1\right) }$ and
applying an operator $I_{2}\otimes H_{d}^{\otimes n}$ on $R$, where $H_{d}$
is the Hadamard gate. The states $|\varphi _{j}\rangle $ are eigenstates of $%
\widetilde{H}$ with eigenvalues of $\alpha $ and degeneracy of $N$.
Therefore, the reference state $|\Psi _{0}\rangle $ has an eigenvalue $%
E_{0}=\alpha $.

We then make a guess on the range of the transition frequencies, $\left[
\omega _{\min }\text{, }\omega _{\max }\right] $, between the reference
state $|\Psi _{0}\rangle $ and the excited states $|\Psi _{j}\rangle
=|1\rangle |\lambda _{j}\rangle $, of $R$, where $|\lambda _{j}\rangle $ ($%
j=1,2,\ldots ,N$) are the $j$-th energy eigenstates of the system with
eigenvalues $E_{j}$. As in Ref.~\cite{wan}, this frequency range is
discretized into $m$ intervals, where each interval has a width of $\Delta
\omega =\left( \omega _{\max }-\omega _{\min }\right) /m$. The center
frequencies are $\omega _{k}=\omega _{\min }+(k+1/2)\Delta \omega ,$ $%
k=0\ldots, m-1$, and these frequency points form a frequency set.

We set the frequency of the probe qubit to be $\omega _{k}$, and let the
entire system evolve with Hamiltonian $H$ for time $\tau $. Then read out
the state of the probe qubit by performing a measurement on the probe qubit
in basis of $|0\rangle $ and $|1\rangle $, which represent the ground and
excited states of the probe, respectively. We repeat the whole procedure
many times to obtain the decay probability of the probe qubit. Then set the
probe qubit in another frequency and repeat the above procedure until run
over all the frequency points in the frequency set. Once we observe a decay
of the probe qubit, it indicates that an excitation between the reference
state and an excited state of the register $R$ occurs and the last $n$
qubits of $R$ collapse to an eigenstate of the system.
\begin{figure}[tbp]
\includegraphics[width=0.8\columnwidth, clip]{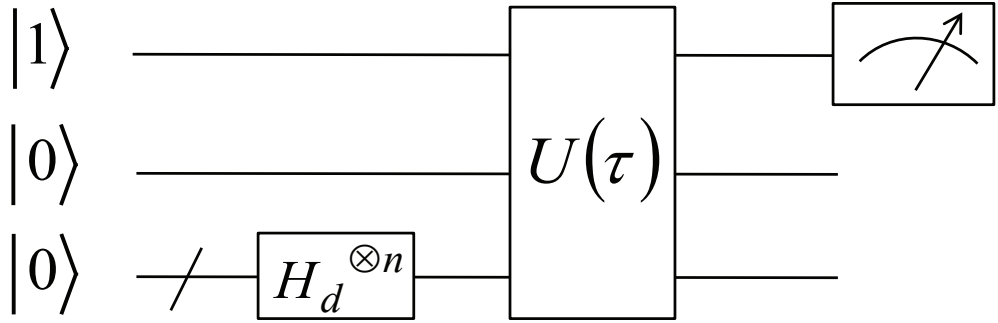}
\caption{Quantum circuit for obtaining the energy spectrum of a physical
system. The first line represents a probe qubit. $H_{d}$ represents the
Hadamard gate, and $U(\protect\tau )$ is a time evolution operator driven by
a Hamiltonian given in Eq.~($1$). The last $n$ qubits represent the system
whose spectrum is to be obtained.}
\end{figure}

The procedure of the algorithm is summarized as follows: $\left( i\right) $
prepare a probe qubit in its excited state $|1\rangle $ and a ($n+1$)-qubit
quantum register $R$ in state $|0\rangle ^{\otimes \left( n+1\right) }$; $%
\left( ii\right) $ apply operator $I_{2}\otimes H_{d}^{\otimes n}$ on the
register $R$, then $R$ is prepared in the reference state $|\Psi _{0}\rangle
$; $\left( iii\right) $ implement the time evolution operator $U(\tau )=\exp
\left( -iH\tau \right) $; $\left( iv\right) $ read out the state of the
probe qubit in basis of $|0\rangle $ and\ $|1\rangle $; $\left( v\right) $
perform steps $\left( i\right) $ -- $\left( iv\right) $ a number of times to
obtain the decay probability of the probe qubit; $\left( vi\right) $ repeat
steps $\left( i\right) $--$\left( v\right) $ for different frequencies of
the probe qubit. The quantum circuit for steps $\left( i\right) $ -- $\left(
iv\right) $ is shown in Fig.~$1$.

In this algorithm, we must implement the time evolution operator $%
U(\tau)=\exp \left( -iH\tau \right) $. This operator can be implemented
based on the Trotter-Suzuki formula~\cite{nc}:
\begin{equation}
U\!(\tau)\!\!=\!\!\left[ e^{-i\frac{1}{2}\omega \sigma _{z}\tau /L}e^{-i%
\widetilde{H}\tau /L}e^{-i\left( c\sigma _{x}\otimes A\right) \tau /L} %
\right]^{L}\!+\!O\!\left( \frac{1}{L}\right),
\end{equation}%
where $L$ does not depend on the size of the problem. $L$ can be made
sufficiently large such that the error is bounded by some threshold~\cite{sl}%
. In Eq.~($5$), the first term of $U(\tau )$, $e^{-i\frac{1}{2}\omega \sigma
_{z}\tau /L}$ is diagonal and can be implemented efficiently on a quantum
computer. And the second term can be treated as a controlled-$U_{S}$
operation, where $U_{S}=e^{-iH_{S}\tau /L}$. $H_{S}$ is a Hamiltonian that
involves two-body interaction, and can be simulated efficiently on a quantum
computer. Therefore the second term of $U(\tau )$ can also be implemented
efficiently. For the third term, $e^{-i\left( c\sigma _{x}\otimes A\right)
\tau /L}$, the Hamiltonian $c\sigma _{x}\otimes A$ involves many-body
interaction. In Ref.~\cite{bravyi}, it was shown that a many-body
interaction Hamiltonian can be simulated efficiently by a Hamiltonian with
two-body interactions. The Hamiltonian $c\sigma _{x}\otimes A$ is equal to $%
c\left( H_{d}\sigma _{z}H_{d}\right) \otimes \left( H_{d}\sigma
_{z}H_{d}\right) \otimes \left[ H_{d}\left(
\begin{array}{cc}
\sqrt{2} & 0 \\
0 & 0%
\end{array}%
\right) H_{d}\right] ^{\otimes n}$, and the unitary operator $e^{-i\left(
c\sigma _{x}\otimes A\right) \tau /L}$ can be written as $H_{d}^{\otimes
\left( n+2\right) }\exp \left[ -ic\tau \sigma _{z}\otimes \sigma _{z}\otimes
\left(
\begin{array}{cc}
\sqrt{2} & 0 \\
0 & 0%
\end{array}%
\right) ^{\otimes n}\right] H_{d}^{\otimes \left( n+2\right) }$. It can be
implemented using the circuit shown in Fig.~$2$. In the circuit, the $\left(
n+1\right) $ and $n$-qubit controlled unitary operators can be implemented
efficiently with $O\left( n^{2}\right) $ elementary gates~\cite{barenco}.
Therefore the unitary operator $U(\tau )$ can be implemented efficiently on
a quantum computer.
\begin{figure}[tbp]
\includegraphics[width=0.8\columnwidth, clip]{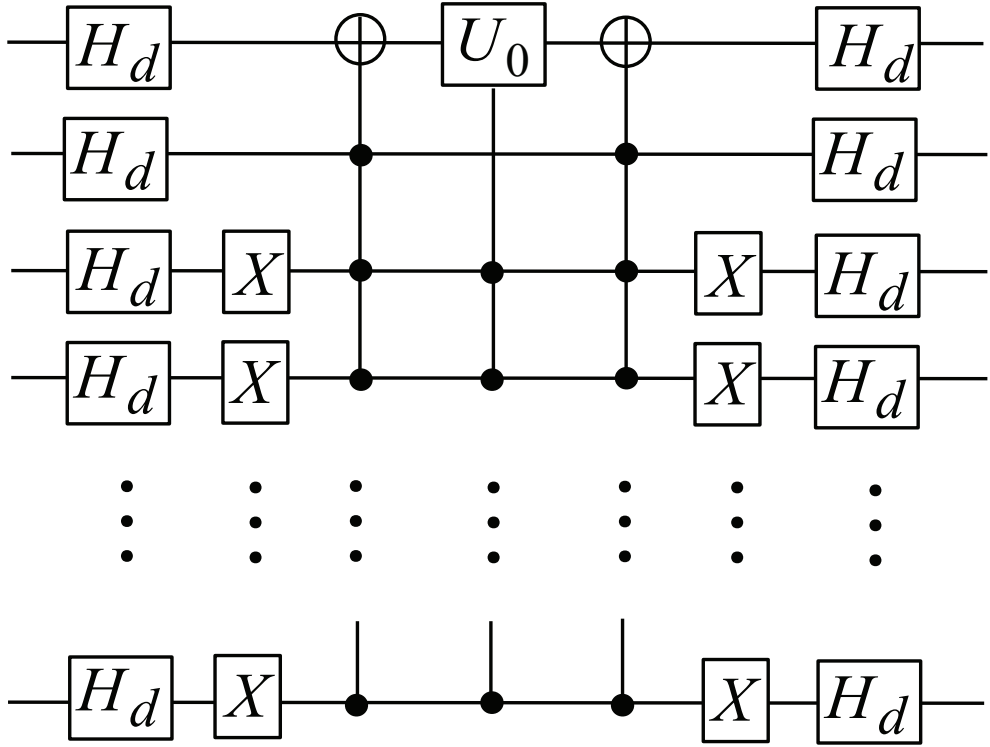}
\caption{Quantum circuit for implementing the unitary operator $e^{-i(c%
\protect\sigma _{x}\otimes A)\protect\tau /L}$, where operator $A$ is given
in Eq.~($3$), $\protect\sigma _{z}$ is the Pauli matrices and $U_{0}=e^{i%
\protect\sqrt{2^{n}}c\protect\tau /L\protect\sigma _{z}}$.}
\end{figure}

\section{Efficiency of the algorithm}

As discussed in Ref.~\cite{wan}, the efficiency of the algorithm is defined
as the number of times that the circuit in Fig.~$1$ must be run to obtain
the decay probability of the probe qubit, $P_{\text{decay}}$. And it must be
at least proportional to $1/P_{\text{decay}}$. In our algorithm, consider
the excitation from the reference state $|\Psi _{0}\rangle $ to the $j$-th
excited state of $R$, $|\Psi _{j}\rangle $, with the probe qubit frequency
being set to $\omega _{k}$, the decay probability of the probe qubit is%
\begin{equation}
P_{\text{decay}}\!\!=\!\!\sin ^{2}\!\left( \!\frac{\Omega _{0j}\tau }{2}%
\!\right) \!\!\frac{Q_{0j}^{2}}{Q_{0j}^{2}\!+\!\left(
\!E_{j}\!-\!E_{0}\!-\!\omega _{k}\!\right) ^{2}},\text{ }j\!=\!1,2,\ldots ,N
\end{equation}%
where $Q_{0j}=2c|\langle \Psi _{j}|A|\Psi _{0}\rangle |$, and $\Omega _{0j}=%
\sqrt{Q_{0j}^{2}+\left( E_{j}-E_{0}-\omega _{k}\right) ^{2}}$. Eq.~($6$)
describes the Rabi oscillation process in which the probe qubit exchanges an
excitation with the register $R$. The probe decays from its excited state to
the ground state, while $R$ is transferred from the reference state $|\Psi
_{0}\rangle $ to the excited state $|\Psi _{j}\rangle =|1\rangle |\lambda
_{j}\rangle $, and the system collapses to its eigenstate $|\lambda
_{j}\rangle $.

The excited states $|\Psi _{j}\rangle $ of the register $R$ can be spanned
in computational basis as%
\begin{equation}
|\Psi _{j}\rangle =\sum_{k=1}^{N}d_{jk}|\mu _{k}\rangle
=\sum_{k=1}^{N}d_{jk}|1\rangle |k-1\rangle .
\end{equation}%
Then $Q_{0j}$ can be written as
\begin{eqnarray}
Q_{0j} &=&2c|\langle \Psi _{j}|A|\Psi _{0}\rangle |  \notag \\
&=&2c\sum_{k=1}^{N}\sum_{l=1}^{N}\frac{1}{\sqrt{N}}d_{jk}|\langle 1|\langle
k-1|A|\varphi _{l}\rangle |  \notag \\
&=&2c\sum_{k=1}^{N}d_{jk}.
\end{eqnarray}%
From Eq.~($8$), we can see that the decay probability and thus the
efficiency of the algorithm, depends on the coupling strength $c$, the
evolution time $\tau $ and the term $\sum_{k=1}^{N}d_{jk}$, which is the
summation of the vector elements of the $j$-th eigenstate of the system. As
we have discussed in Ref.~\cite{wan}, as long as the number of
\textquotedblleft energy levels of interest\textquotedblright\ is
polynomially large, the complexity of the algorithm grows polynomially with
the size of the system.

It should be pointed out that the coupling strength $c$ is small ($c\ll
\omega $) and the effect of the perturbation of the probe qubit to the
register $R$ is sufficiently weak. In this case, its effect on $R$ can be
calculated to a first approximation, by ignoring all the other energy levels
of the register $R$.

The coupling between the reference state and all the other energy levels
except the one that resonant with the probe, contributes to the decay
probability of the probe qubit, therefore introduces an error, $P_{\text{%
decay}}^{\text{err}}$, in $P_{\text{decay}}$. For a system with discrete
energy levels, when there is no energy level that has exponentially large
degeneracy, the error $P_{\text{decay}}^{\text{err}}$ can be constrained to
be very small because $c$ can be set polynomially small. It was shown that
the simulation of any Hamiltonian may be performed linearly in the scaled
time $\tau $~\cite{berry}. And in our algorithm we have $c\tau \sim 1$.
Therefore, in this case, the algorithm can be run in finite time $\tau $.
For a system with exponentially large number of degenerated states, we may
not find a coupling coefficient $c$ that is polynomially small such that the
evolution time is finite~\cite{whf1}.

We consider the case where the transition between states $|\Psi _{0}\rangle $
and $|\Psi _{1}\rangle $ resonates with the probe qubit with frequency of $%
\omega $, such that $E_{1}-E_{0}=\omega $. The error $P_{\text{decay}}^{%
\text{err}}$ can be calculated as follows
\begin{eqnarray}
P_{\text{decay}}^{\text{err}} &=&\sum_{j=2}^{N}\sin ^{2}\left( \frac{\Omega
_{0j}\tau }{2}\right) \frac{Q_{0j}^{2}}{Q_{0j}^{2}+\left( E_{j}-E_{0}-\omega
\right) ^{2}}  \notag \\
&<&\sum_{j=2}^{N}\frac{Q_{0j}^{2}}{\left( E_{j}-E_{1}\right) {}^{2}}  \notag
\\
&=&4c^{2}\sum_{j=2}^{N}\frac{\left( \sum_{k=1}^{N}d_{jk}\right) ^{2}}{\left(
E_{j}-E_{1}\right) {}^{2}}.
\end{eqnarray}%
For a finite system, if the ground state is not degenerate and the term $%
\sum_{j=2}^{N}\frac{\left( \sum_{k=1}^{N}d_{jk}\right) ^{2}}{\left(
E_{j}-E_{1}\right) {}^{2}}$ is bounded by a finite number $M$. The term $%
4c^{2}M$ can be small by setting $c$ small. In this case, the algorithm can
be completed in finite time. That is, for a finite system, a sufficiently
small $c$ and a finite evolution time $\tau $ exist.

\section{Example: obtaining the energy spectrum of the water molecule}

In the following, using the water molecule as an example, we simulate the
algorithm for obtaining the energy spectrum of a system that we have no
information about its eigenstates.

The Hamiltonian of the water molecule is the same as shown in Ref~\cite{wan}%
. Considering the C$_{2V}$ and $^{1}A_{1}$ symmetries, the Hartree-Fock wave
function for the ground state of the water molecule is $%
(1a_{1})^{2}(2a_{1})^{2}(1b_{2})^{2}(3a_{1})^{2}(1b_{1})^{2}$. Using the STO-%
$3$G basis set~\cite{sbo} and freezing the first two $a_{1}$ orbitals, a
model space with $^{1}A_{1}$ symmetry that includes the $%
3a_{1},4a_{1},1b_{1} $ and $1b_{2}$ orbitals is constructed by considering
only single and double excitations to the external space. For simplicity, we
remove two of the highest excitations, then the dimension of the state space
of the water molecule is $16$. Therefore four qubits are required to
simulate the water molecule in this calculation.
\begin{figure}[tbp]
\includegraphics[width=0.88\columnwidth, clip]{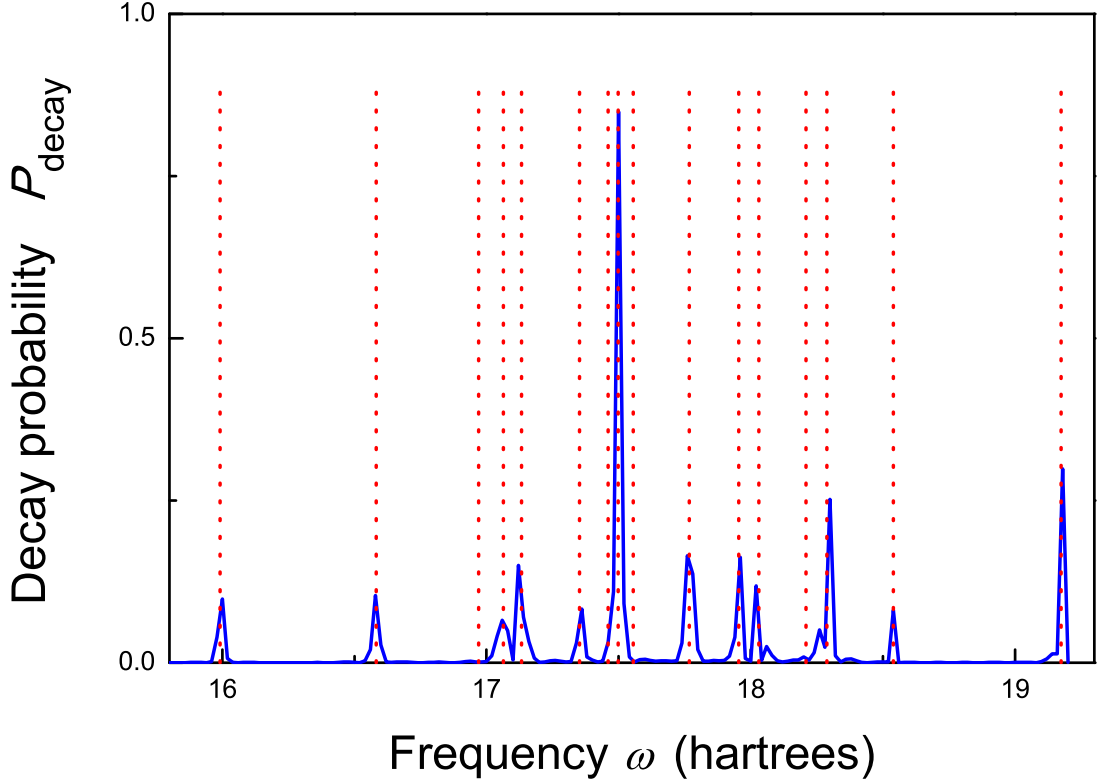}
\caption{(Color online)~Transition frequency spectrum between the reference
state, $|\Psi _{0}\rangle $, and the $16$ eigenstates of the water molecule.
The blue solid curve represents the decay probability of the probe qubit at
different frequencies with the coupling coefficient $c=0.002$ and the
evolution time $\protect\tau =1200$. The frequency for the probe qubit is
set in the range $\protect\omega \in \left[ 15.8\text{, }19.2\right] $, and
is divided into $170$ equal intervals. The red dotted vertical lines
represent the known transition frequencies between the reference state and
all the $16$ eigenstates of the water molecule.}
\end{figure}

We set the reference energy $E_{0}=\alpha =-100$, and vary the frequency of
the probe qubit in the range $\omega \in \left[ 15.8\text{, }19.2\right] $,
which is divided into $170$ equal intervals. The coupling strength and the
evolution time are set as $c=0.002$ and $\tau =1200$~(here energies and time
are measured in units of Hartree and Hartree$^{-1}$, respectively). Then we
run the algorithm and obtain the spectrum of the transition frequencies
between the reference state and the eigenstates of the water molecule. The
results are shown in figure~$1$. From the figure we can see that most of the
spectrum obtained using our algorithm are in good agreement with the known
transition frequency spectrum~(in red) of the water molecule, except that
four peaks~(the $3$rd, $7$-th, $9$-th, and $13$-th) are missing.

Some missing peaks can be found by only increasing the density of the
frequency points in certain frequency range. We vary the frequency of the
probe qubit in the range $\omega \in \left[ 15.8\text{, }17.0\right] $ and
divide this frequency range into $240$ equal intervals, run the algorithm.
The results are shown in figure~$2(a)$. We can see that the $3$rd peak at $%
\omega =16.9705$ is visible now. We then vary the frequency of the probe in
the range $\omega \in \left[ 17.2\text{, }18.0\right] $, which is divided
into $160$ equal intervals, run the algorithm, from the results shown in
figure~$2(b)$, we can see that the $9$-th peak at $\omega =17.5552$ appears.
We vary the frequency of the probe in the range $\omega \in \left[ 18.0\text{%
, }19.2\right] $, which is also divided into $240$ equal intervals, and set $%
c=0.001$ and $\tau =2400$. From the results shown in figure~$2(c)$, we can see
that the $13$-th peak at $\omega =18.2082$ is clearly visible now.

In the case where the term $\sum_{k=1}^{N}d_{jk}$ is small~(then the decay
probability is small), increasing the evolution time $\tau $ can increase
the height of the peak. For the $7$-th eigenstate of the system, we have $%
\sum_{k=1}^{N}d_{jk}=0.0305153$, which is a small number. We vary the
frequency of the probe in the range $\omega \in \left[ 17.4\text{, }17.6%
\right] $, which is divided into $400$ equal intervals, set $c=0.001$ and $%
\tau =20000$, then run the algorithm. The results are shown in figure~$2(d)$.
We can see that the $7$-th peak at $\omega =17.4594$ appears now.
\begin{figure}[tbp]
\begin{minipage}{0.48\linewidth}
  \centerline{\includegraphics[width=0.98\columnwidth, clip]{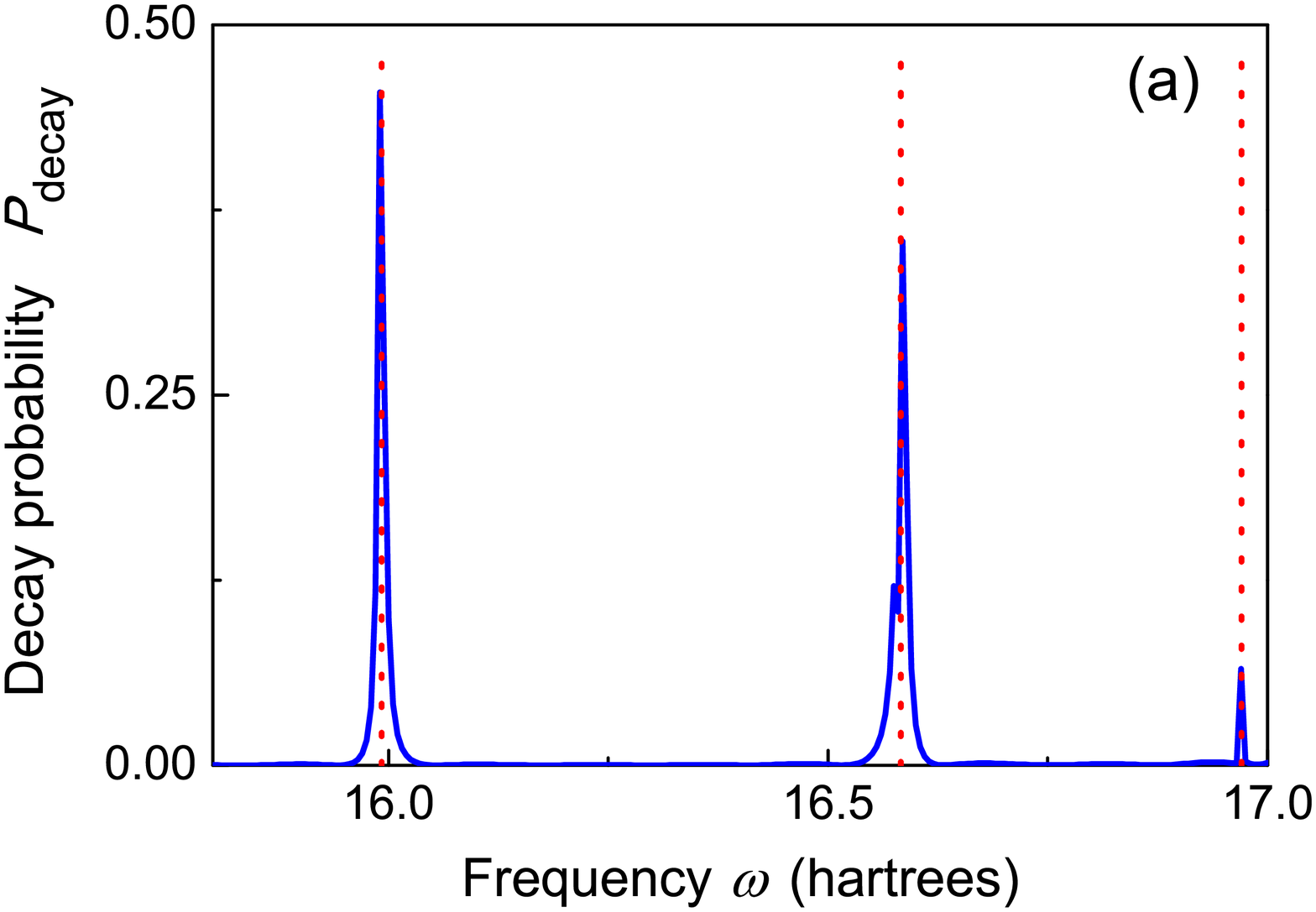}}
\end{minipage}
\hfill
\begin{minipage}{0.48\linewidth}
  \centerline{\includegraphics[width=0.98\columnwidth, clip]{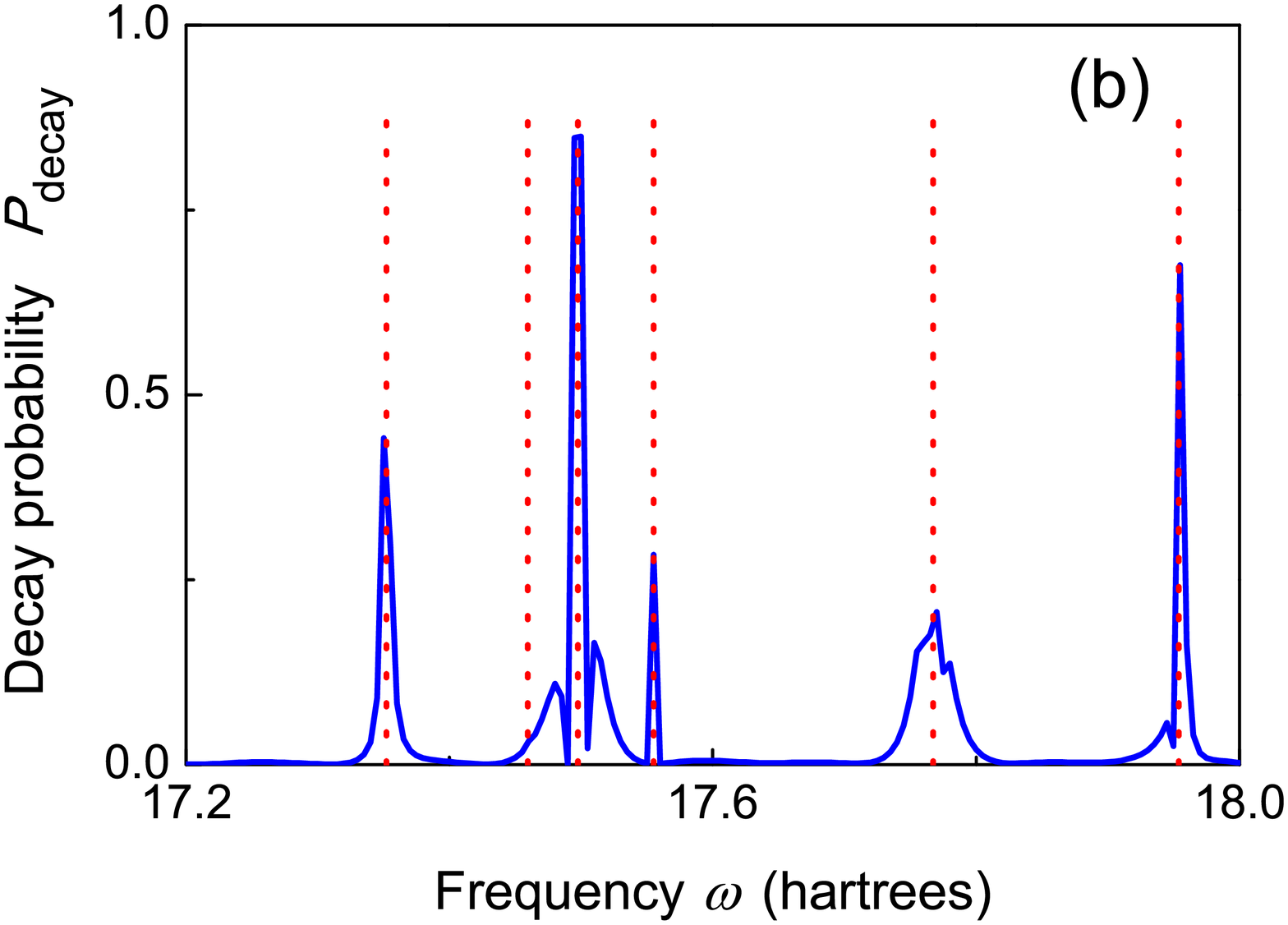}}
\end{minipage}
\vfill
\begin{minipage}{0.48\linewidth}
  \centerline{\includegraphics[width=0.98\columnwidth, clip]{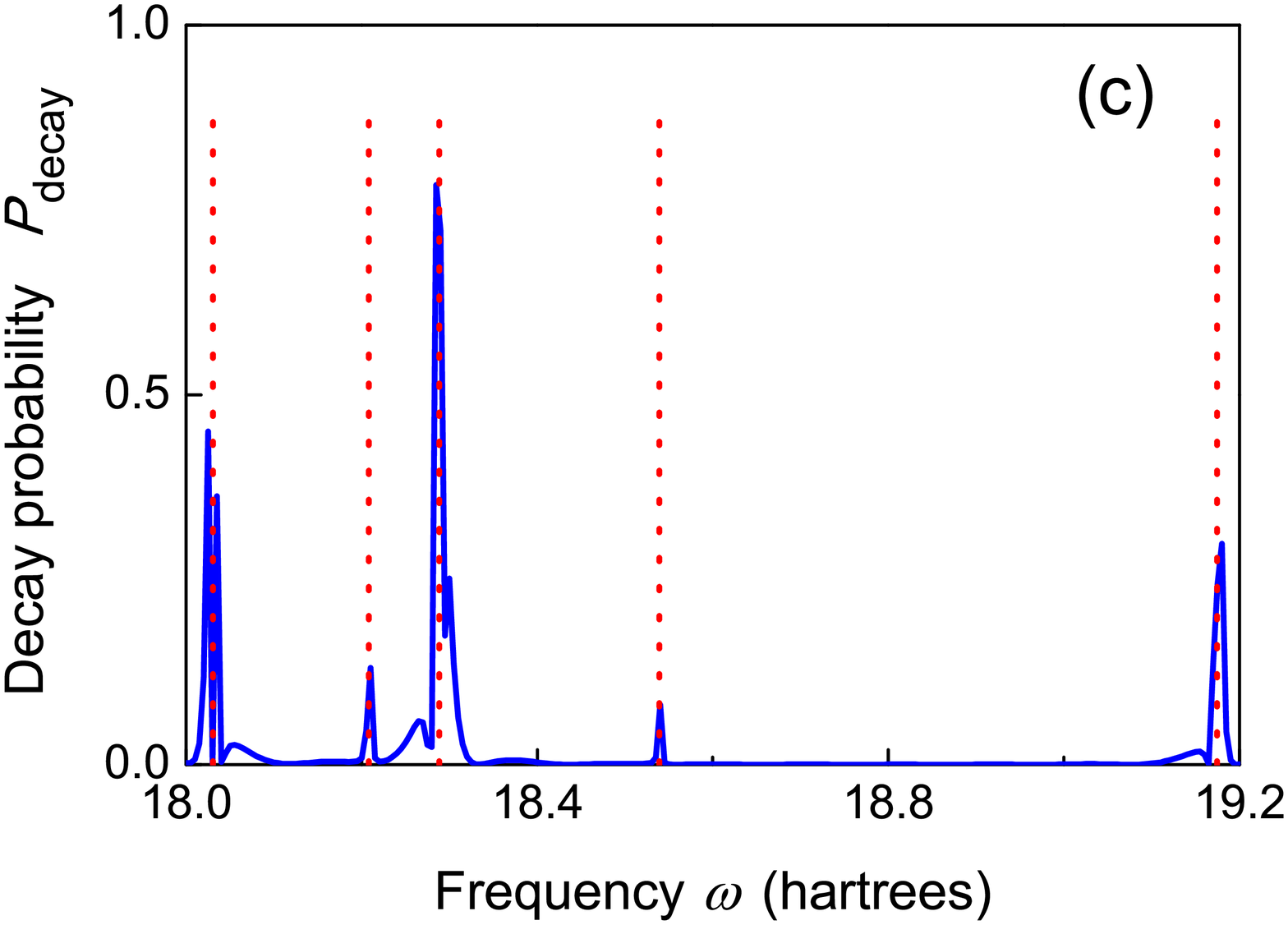}}
\end{minipage}
\hfill
\begin{minipage}{0.48\linewidth}
  \centerline{\includegraphics[width=0.98\columnwidth, clip]{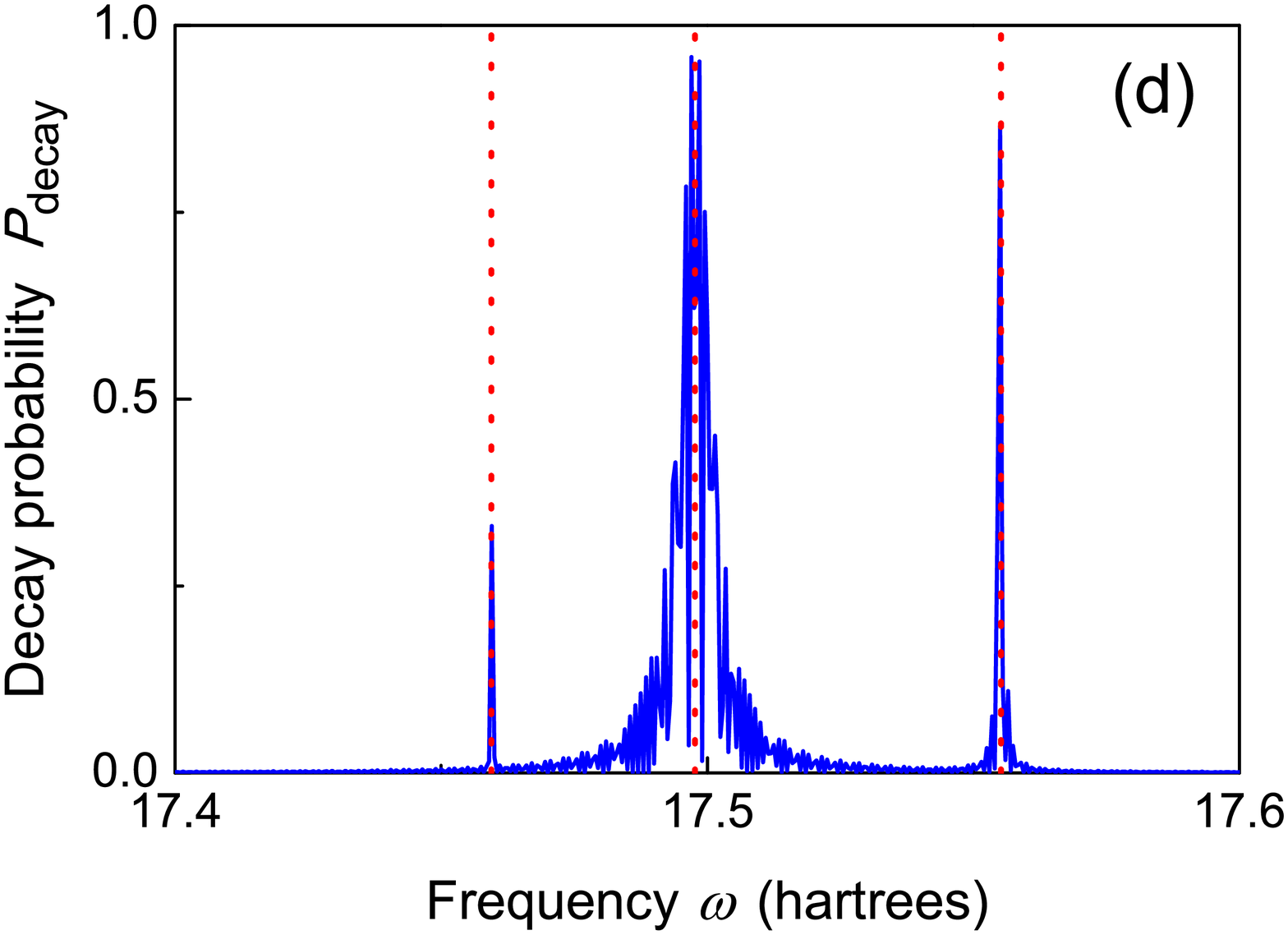}}
\end{minipage}
\caption{(Color online)~Same as in Fig.~$3$, except that in $(a)$ and $(b)$,
$c=0.002$ and $\protect\tau =1200$. In $(a)$ the frequency of the probe
qubit $\protect\omega $ is set in the range $\protect\omega \in \left[ 15.8%
\text{, }17.0\right] $ and is divided into $240$ equal intervals; in $(b)$ $%
\protect\omega \in \left[ 17.2\text{, }18.0\right] $ and is divided into $%
160 $ equal intervals. In $(c)$, $c=0.001$, $\protect\tau =2400$, $\protect%
\omega \in \left[ 18.0\text{, }19.2\right] $ and is divided into $240$ equal
intervals. In $(d)$, $c=0.001$, $\protect\tau =20000$, $\protect\omega \in %
\left[ 17.4\text{, }17.6\right] $ and is divided into $400$ equal intervals.}
\end{figure}

\section{Discussion}

As we have demonstrated in the example for obtaining the energy spectrum of water molecule in the supplementary material, by varying the coupling strength $c$%
, the evolution time $\tau $ and increasing the density of the frequency
points, one can obtain the entire energy spectrum and the corresponding
eigenstates of a system. As we have discussed in Ref.~\cite{wan}, the
accuracy of the algorithm is defined by the parameters $c$ and $\tau $. We
need to set $c$ to be small so that the system-probe coupling is weak, and
the evolution time $\tau $ to be large. The size of the frequency intervals $%
\Delta \omega $ is set by the choice of $c$ and $\tau $: $\Delta \omega $
should be smaller than the width of the peaks in order to avoid missing some
of the peaks.

It should be pointed out that our algorithm cannot be used for obtaining the
eigenenergies of eigenstates that have anti-symmetric symmetry. As shown in
Eq.~($8$), the term $Q_{0j}$ is zero therefore the decay probability of the
probe is zero. By using some other techniques, such as group theory, one can
determine the anti-symmetric state and using the PEA to obtain its
eigenenergy. Also, in our algorithm, one cannot tell whether a given energy
level is degenerate or not. It is also difficult to separate the
near-degenerate states. In these cases, by applying our algorithm, once the
probe qubit collapses to its ground state, the system is in a superposition
of the degenerated eigenstates of the system. One can use this state as the
input for the PEA to resolve the eigenenergy and the corresponding
eigenstates.

We now compare this algorithm with the algorithm we proposed in Ref.~\cite%
{wan}. In the previous algorithm, we prepare the system in an initial state
that is close to one of its eigenstates, and construct an excitation
operator that transfers the initial state to another state of the system. By
coupling with a probe qubit, the system is evolved to the desired
eigenstates of the system. The form of the excitation operator depends on
the guess state and the part of the energy spectrum that is of interest.

In this algorithm, one ancilla qubit is added to the register of the system
to construct a quantum register $R$, and $R$ is coupled to a probe qubit.
Here $R$ can be considered as an \textquotedblleft artificial
system\textquotedblright\ which plays the same role as \textquotedblleft the
system\textquotedblright\ in the previous work~\cite{wan}.

The Hamiltonian of the register $R$ given in Eq.~($2$) can be written as: $%
\widetilde{H}=\alpha |0\rangle \langle 0|\otimes
I_{N}+\sum_{j=1}^{N}E_{j}|1\rangle \langle 1|\otimes |\lambda _{j}\rangle
\langle \lambda _{j}|=\alpha |0\rangle \langle 0|\otimes
I_{N}+\sum_{j=1}^{N}E_{j}|\Psi _{j}\rangle \langle \Psi _{j}|$ ($E_{j}$ are
the eigenenergies of the system). Its ground state is $N$-fold degenerate
and in the form of $|0\rangle |j-1\rangle $~($j=1,2,\ldots ,N$), which are
the eigenstates of the first term of $\widetilde{H}$. And its excited states
are $|\Psi _{j}\rangle =$ $|1\rangle |\lambda _{j}\rangle $~($|\lambda
_{j}\rangle $ are the eigenstates of the physical system), which are the
eigenstates of the second term of $\widetilde{H}$. The register $R$ is
prepared in the reference state $|\Psi _{0}\rangle $ which is the eigenstate
of the first term with eigenvalue $\alpha $. We introduced an excitation
operator $A$ as defined in Eq.~($3$) acting on the register $R$. From the
expansion of $A$, one can see that $A$ contains $N$ terms which provide all
possible excitations between the subspace of $|0\rangle \langle 0|\otimes
I_{N}$ and the subspace of $|1\rangle \langle 1|\otimes H_{S}$. The probe
qubit is coupled to $R$ with interaction operator $c\sigma _{x}\otimes A$,
which transfers the probe from the excited state to its ground state and $R$
from the reference state to a state in the subspace of $|1\rangle \langle
1|\otimes H_{S}$. The overlap of this state with the excited state $|\Psi
_{j}\rangle $ is $\langle \Psi _{j}|A|\Psi _{0}\rangle $ and has been
derived in Eq.~($8$). By employing the operator $A$, $R$ can be evolved to
any of its excited states starting from the initial state $|\Psi _{0}\rangle
$. The probe qubit exhibits a dynamical response only when it is resonant
with a transition between the reference state $|\Psi _{0}\rangle $ and a
state $|\Psi _{j}\rangle $ of $R$. Therefore, when $E_{j}-\alpha =\omega $ ($%
\omega $ is the frequency of the probe qubit), the probe qubit decays to its
ground state with decay probability $P_{\text{decay}}=\sin
^{2}\left(\frac{Q_{0j}\tau }{2}\right) $ while the register $R$ is
transferred to state $|\Psi _{j}\rangle $. Therefore for a finite system, as
long as the term $Q_{0j}$ is not exponentially small, the algorithm can be
run efficiently. Another advantage of employing operator $A$ is that in the
Trotter expansion, the unitary operator related to the interaction operator
can be implemented efficiently as we have shown in Fig.~($2$).

In this algorithm, we introduced a reference state $|\Psi_{0}\rangle $ and
an excitation operator $A$, both do not depend on systems. They are general
and can be applied for any arbitrary physical system. This algorithm
provides a general approach for obtaining the energy spectrum and energy
eigenstates of a physical system without having any information about the
eigenstates of the system.

\begin{acknowledgements}
We thank Sahel Ashhab and L.-A. Wu for helpful discussions. This work was supported by \textquotedblleft the Fundamental Research Funds for the Central Universities\textquotedblright\ of China and the National Nature Science Foundation of China~(Grants No.~11275145 and No.~11305120).
\end{acknowledgements}


\begin{thebibliography}{99}
\bibitem{davidson} E.~R. Davidson, J. Comp. Phys. \textbf{17}, 87~(1975).

\bibitem{abrams} D.~S. Abrams and S. Lloyd, Phys. Rev. Lett. \textbf{83},
5162~(1999).

\bibitem{whf} H. Wang, S. Kais, A. Aspuru-Guzik, and M.~R. Hoffmann, Phys.
Chem. Chem. Phys. \textbf{10}, 5388~(2008).

\bibitem{wan} H. Wang, S. Ashhab, and F. Nori, Phys. Rev. A \textbf{85},
062304~(2012).

\bibitem{nc} M. Nielsen, I. Chuang, \emph{Quantum Computation and Quantum
Information}~(Cambridge Univ. Press, Cambridge 2000).

\bibitem{sl} S. Lloyd, \emph{Science} \textbf{273}, 1073-1078~(1996).

\bibitem{bravyi} S. Bravyi, D.~P. DiVincenzo, D. Loss, and B.~M. Terhal,
Phys. Rev. Lett. \textbf{101}, 070503~(2008).

\bibitem{barenco} A. Barenco, \emph{et al.}, Phys. Rev. A \textbf{52},
3457~(1995).

\bibitem{berry} D.~W. Berry, G. Ahokas, R. Cleve, and B.~C. Sanders, Comm.
in Math. Phys. \textbf{270}, 359~(2007)

\bibitem{whf1} H. Wang, submitted.

\bibitem{sbo} A. Szabo and N. Ostlund, {\normalsize \textit{Modern Quantum
Chemistry: Introduction to advanced Electronic Structure Theory}}%
~(McGraw-Hill, New York, 1989).

\end{thebibliography}
\end{document}